# Parallelization of JOREK-STARWALL for non-linear MHD simulations including resistive walls (Report of the EUROfusion High Level Support Team Projects JORSTAR/JORSTAR2)

S. Mochalskyy, M. Hoelzl, R. Hatzky

## 1. Introduction

Large scale plasma instabilities inside a tokamak can be influenced by the currents flowing in the conducting vessel wall. This involves non linear plasma dynamics and its interaction with the wall current. In order to study this problem the code that solves the magneto-hydrodynamic (MHD) equations, called JOREK [1,2], was coupled [3] with the model for the vacuum region and the resistive conducting structure named STARWALL [4,5]. The JOREK-STARWALL model has been already applied to perform simulations of Vertical Displacement Events (VDEs) [6], Resistive Wall Modes (RWMs) [13], Quiescent H-Mode [6], and vertical kick ELM triggering [7].

At the beginning of the project it was not possible to resolve the realistic wall structure with a large number of finite element triangles due to the huge consumption of memory and wall clock time by STARWALL and the corresponding coupling routine in JOREK. Moreover, both the STARWALL code and the JOREK coupling routine were only partially parallelized via OpenMP. The aim of this project is to implement an MPI parallelization to reduce memory consumption and execution time such that simulations with large resolutions become possible.

The project JORSTAR is concerned with the MPI parallelization of STARWALL (chapter 2). In the project JORSTAR2, the implementation of parallel I/O in JOREK and STARWALL for the STARWALL response matrices, and the parallelization of the JOREK-STARWALL coupling terms inside JOREK are addressed (chapter 3).

## 2. MPI parallelization of the resistive wall code STARWALL

### 2.1. STARWALL code analysis

It was important to determine the most critical data structures and subroutines that consume most of the memory and execution time before starting the implementation of the MPI parallelization. The memory consumption and the execution time for individual subroutines concerning different problem sizes can be controlled by tuning three knobs, which directly influence the problem size (a test case with a closed axisymmetric wall is considered):

- Number of triangles representing the boundary of the JOREK computational domain:
  $ntri\_p = 4*nv*n\_points*2*(n\_R+n\_Z-2)$
- Number of triangles in the wall: $ntri\_w = 2*nwu*nwv$
- Number of sin/cos harmonics: $n\_harm$

We changed the problem size by varying the following parameters independently: (i) $n\_R$ and $n\_Z$ for $ntri\_p$, (ii) $nwu$ and $nwv$ for $ntri\_w$, and (iii) $n\_harm$. A large scale production run should finally correspond to the parameters: $ntri\_p=2*10^5$, $ntri\_w=5*10^5$, $n\_harm=11$.

#### 2.1.1. Memory consumption analysis

Fig. 1 shows the memory consumption of the most important individual subroutines during the scan of the parameter $ntri\_w$ by varying the variables $nwu$ and $nwv$. For this test case we fixed *n_harm*=1, *n_R*=*n_Z*=15, *nv*=32, and *n_points*=10. One can see that three subroutines (*matrix_wp*, *matrix_ww*, and *resistive_wall_response*) are the most memory demanding in this scan. Moreover, if we further scale our problem to a production size run with *nwu*=*nwv*=500 (*ntri_w*=500000) five additional subroutines (*matrix_rw, solver, dsygv, matrix_ew, matrix_pe*) will consume more than 50 GB memory. Therefore, all these subroutines must be parallelized in the final version of the code.

Fig. 2 represents the memory consumption of the same subroutines as it was shown in Fig. 1, however, this time with a parametric scan in the number of triangles within the plasma (*ntri_p*). In this test we kept the following parameters constant *nwu*=*nwv*=110, *n_harm*=1 but changed *n_R*=*n_Z*. The memory consumption increased mainly in three subroutines (*matrix_pp, matrix_wp, and matrix_ep*), which should be parallelized for a production run with *ntri_p*=$2*10^5$.

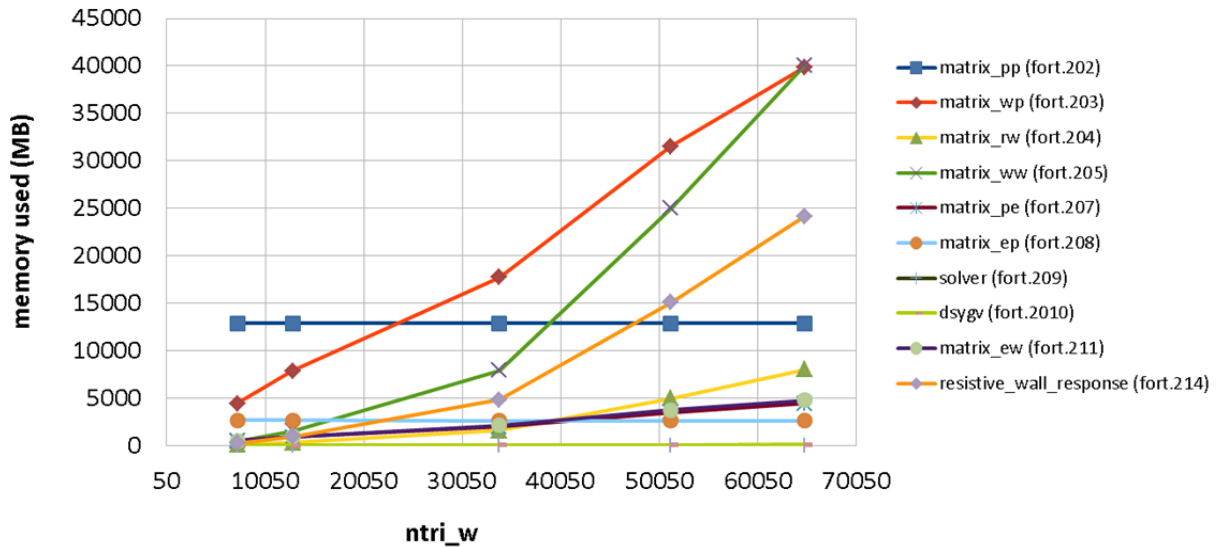

**Fig.** 1 The memory consumption of individual subroutines of the STARWALL code during the scan over the number of the triangles discretizing the wall (*ntri_w*=2\**nwu*\**nwv*).

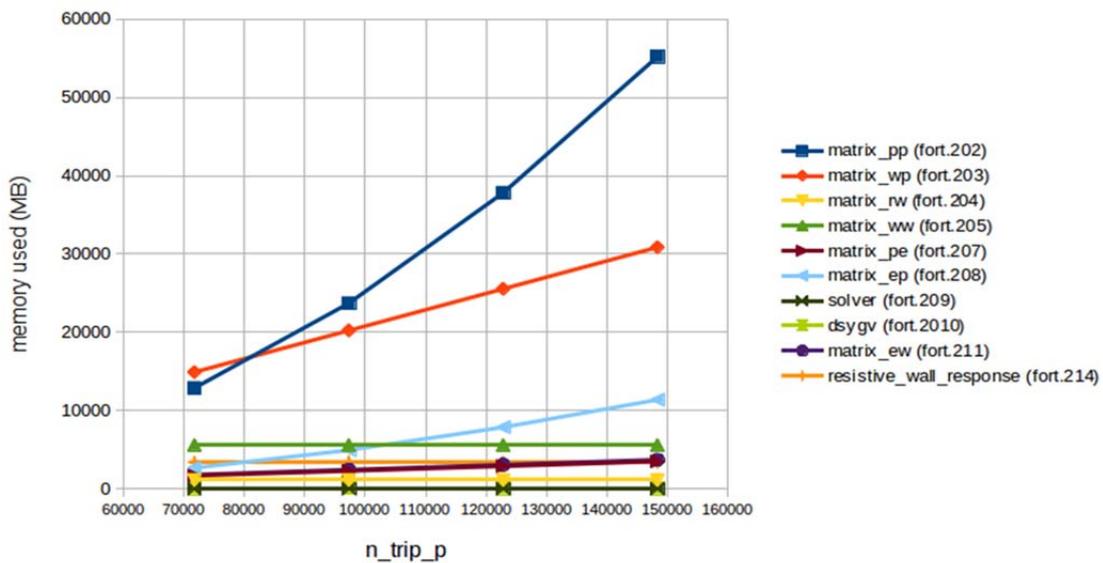

**Fig.** 2 The memory consumption of individual subroutines of the STARWALL code during the scan over the number of the triangles within the plasma (*ntri_p*).

The last parameter tested was the number of sin/cos harmonics (*n_harm*). Fig. 3 shows the memory consumption per subroutine versus *n_harm,* which varies from one to eleven. The value *n_harm*=11 corresponds to a production run. For this testcase we kept the following parameters constant: *nwu*=*nwv*=80, *n_R*=*n_Z*=15. All subroutines stay almost at the same level of memory consumption with only an insignificant growth for some subroutines. In order to prove that the number of sin/cos harmonics will not have a large influence on the memory consumption, whilst the number of triangles is increased, we performed an additional test with *nwu*=*nwv*=110. Indeed, as in the test above, the memory usage did not change much during the *n_harm* scan.

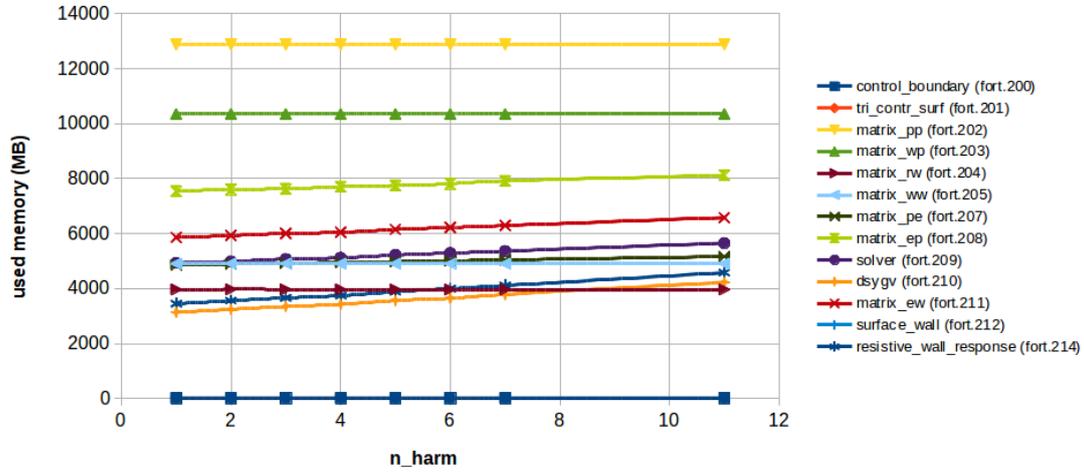

**Fig. 3** The memory consumption of individual subroutines of the STARWALL code during a scan over the number of sin/cos harmonics.

STARWALL uses six subroutines (*dpotrf*, *dpotrs*, *dgemm*, *dsygv*, *dgetrf*, *dgetri*) from the linear algebra package LAPACK that is part of Intel the MKL library. It was important to check both, the size of the input matrices of these subroutines and the additional memory allocation inside the subroutines in order to determine if we should also replace these sequential subroutines by their parallel analogues. A dedicated script was developed for this propose, which measures the time spent executing the LAPACK subroutines and their memory consumption. It was found that only the *dsygv* LAPACK subroutine requires additional allocation of memory, which however, is negligible (~50–100 MB). Finally, the size of the input matrices for the production will range between 20 GB and few TB. Therefore, all LAPACK subroutines must be replaced by their parallel versions from other libraries like ScaLAPACK in order to distribute the input/output matrices, and hence reduce the size of the local sub-matrices.

Summarizing our tests, the complete STARWALL code must be adapted in order to distribute the memory consumption. We estimated that the production run will require about six to seven TB of physical memory that can be allocated by using about 100 computing nodes on the IFERC-CSC HELIOS computer.

### 2.1.2. Computational time analysis

The memory analysis has already shown the necessity of a complete domain decomposition of the whole code. Additionally, it was also important to determine the wall clock time for the production run and find the hot spots in the code. Fig. 4 shows the STARWALL execution time for different amounts of triangles in the wall and within the plasma (red and green lines). For a large scale production simulation on a single CPU the wall clock time would be in the range of a year.

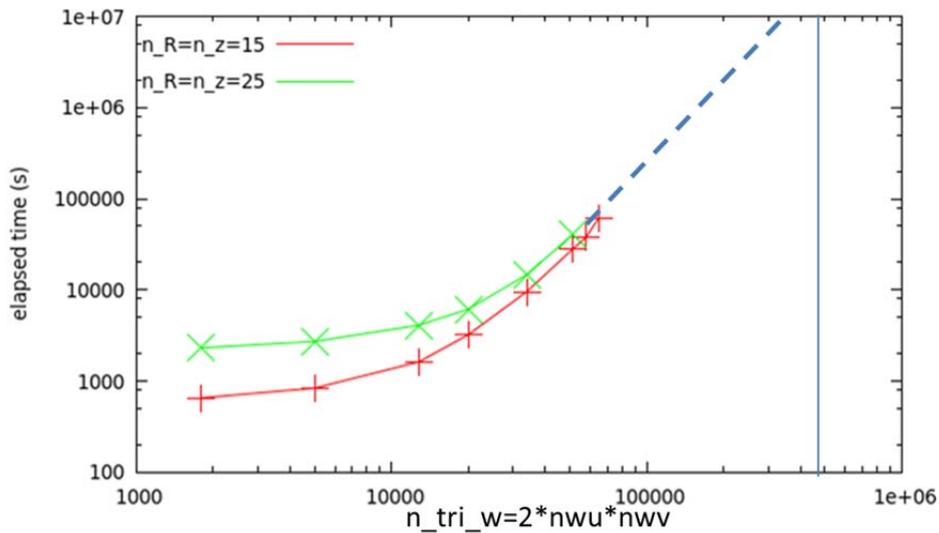

**Fig.** 4 The wall clock time versus the number of triangles in the wall (ntri_w) for different numbers of triangles within the plasma: $n\_R=n\_Z=15$ shown as red line, $n\_R=n\_Z=25$ shown as green line. The solid blue line shows the targeted numbers of triangles for a production run, while the dashed blue line presents the extrapolated scaling.

The next step was to determine the most time consuming subroutines in the code. This analysis was performed by means of the Allinea Forge profiling package. Depending on the problem size different subroutines contribute to a different percentage of the total execution time. However, among all subroutines, one (*dsygv*) consumes in all cases more than 40% of the total wall clock time. For the largest problem size we could run, the percentage was > 70%. Hence, this subroutine became the first candidate for parallelization effort and improvement.

### 2.1.3. OpenMP parallelization analysis

STARWALL is partially parallelized by means of OpenMP directives. Its parallelization efficiency is shown in Fig. 5. The wall clock time decreases by a factor of 1.4 when 16 threads are involved in comparison to the sequential run. Such poor performance can be explained by Amdahl's law, which shows the maximal possible speed-up of a program only partially parallelized. According to this law the maximal speed-up factor we can expect is around two. For this estimate we have taken into account that all LAPACK routines are sequential. With this assumption the sequential parts of STARWALL add up to about 45 percent of the total execution time.

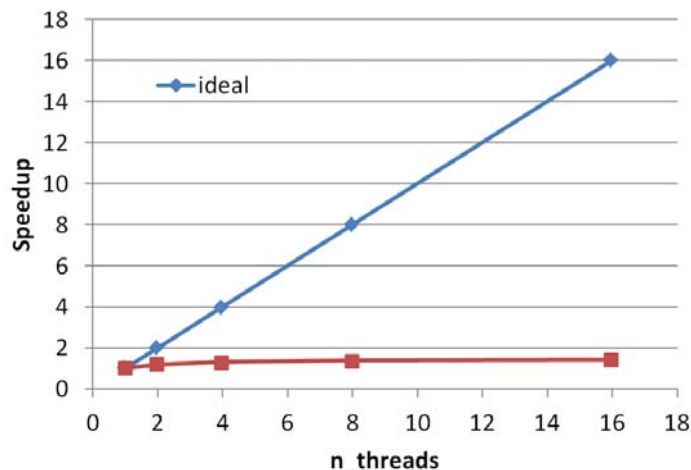

**Fig. 5** Speed-up of the code versus number of OpenMP threads.

In order to confirm poor OpenMP parallelization scalability our model was checked via the Intel Vtune performance profiler. The basic hot spots analysis is presented in

Fig. 6. One can see that for most of the time only one thread is performing calculations (brown color), while the other 15 threads stay idle, as expected. Such results confirm the necessity of a replacement of all sequential LAPACK subroutines with their parallel analogues.

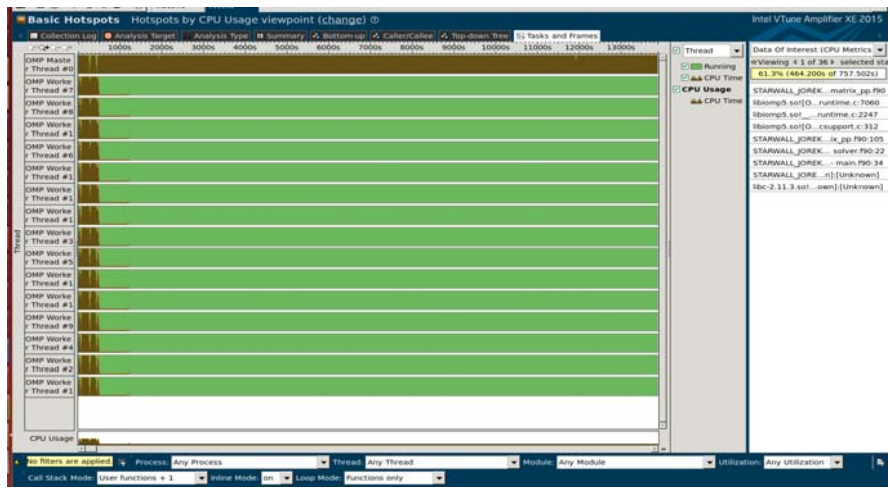

**Fig. 6** Basic Hotspots analysis from the Intel Vtune amplifier using 16 OpenMP threads. Brown color shows the working status of the process, while green color corresponds to the idle state.

### 2.1.4. LAPACK subroutines

As it was discussed earlier the code spends most of the computational time in the execution of the LAPACK subroutines. In this subsection we summarize all LAPACK subroutines which are used in STARWALL:

- *dpotrf* – computes the lower-upper (LU) factorization of a tridiagonal matrix;
- *dpotrs* – solves a system of linear equations with a Cholesky factored symmetric positive defined matrix;
- *dgemm* – computes a matrix-matrix product for general matrices;
- *dsygv* – computes all eigenvalues and corresponding eigenvectors of a real generalized symmetric definite eigenproblem;
- *dgetrf* – computes the LU factorization of a general matrix;
- *dgetri* – computes the inverse of the LU factored general matrix.

### 2.1.5. Bug check

Before starting the optimization and parallelization the code was checked for correctness. The run time debugging was performed with two different compilers: *Lahey* and *Intel*. Afterwards the source code was also analyzed by the *Forcheck* static analyzer.

Three uninitialized variables were found that could produce unexpected behavior of the code:

1) In file solver.f90: $nd\_w=$**ncoil**$+npot\_w$
2) In file matrix_ec.f90: alv=pi2***fnv**
3) In file resistive_wall_respones.f90: **ntri_c**

These problems were reported to the project coordinator and resolved afterwards.

The code was running mainly on a LINUX cluster called *TOK-P,* which is located at RZG, Garching. During parallel simulations a bug was detected in the standard input (*stdin*) system of this cluster. Within the default configuration only the process with *rank*=0 reads data from the *stdin*. Adding the flag *'-s all'* to *mpirun* should allow all processes being involved in the computation to read data from standard input. However, this flag was working only on a single node with all MPI tasks pinned. For tests with two or more nodes the code got stuck at the *stdin* reading. The same tests were performed on *HELIOS* using the same compiler and compile flags. In this case the *std* reading worked properly. This bug was reported to the support team of the TOK-P cluster at RZG. The problem was avoided by reading the input only on task 0 and communicating it to the other tasks.

## 2.2. *MPI parallelization*

### 2.2.1. **Parallelization of the eigenvalue solver**

The LAPACK subroutine used for the calculation of the eigenvalues and the corresponding eigenvectors got the priority for parallelization. This subroutine consumes more than 70% of the total STARWALL execution time and uses two large matrices as input parameters. The subroutine is called *dsygv* and a more detailed description can be found in Ref. [8]. This subroutine was replaced by its parallel version *PDSYGVX* from the ScaLAPACK library that includes subroutines for linear algebra computation on distributed memory computers supporting MPI [8].

*The PDSYGVX* subroutine includes 34 input/output parameters by means of which the user can specify: the eigenvalue problem type to be solved, which eigenvalues and eigenvectors must be computed, the calculation precision, etc. Prior the calculation all global matrices must be distributed on process grid using a so called block-cycling scheme [8].

In order to test the correctness of the implementation of the *PDSYGVX* subroutine the calculated eigenvalues and the eigenvectors were compared with the results from the original (sequential) subroutine *dsygv.* Fig. 7 shows the calculated eigenvalues from both the *dsygv* (red points) and the *PDSYGVX* (green points) subroutines. In the case of the ScaLAPACK subroutines 16 MPI processes distributed over 16 computational nodes (1 per node) were used. A very good agreement was found for different problem sizes.

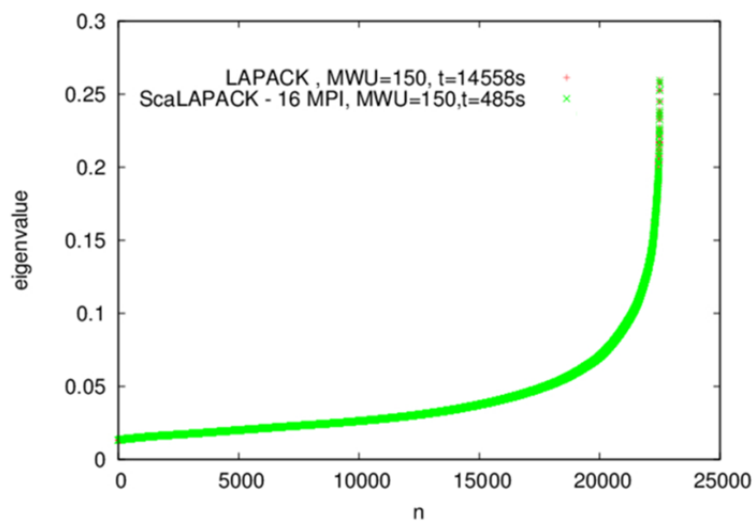

**Fig. 7** Eigenvalues from the sequential LAPACK *dsygv* (red points) and the parallel ScaLAPACK *PDSYGVX* (green points) subroutine.

In spite of the perfect agreement of the eigenvalues the calculated eigenvectors are somehow unpredictable. For some problem sizes they are identical between the *dsygv* and *PDSYGVX* subroutine. In other cases some eigenvectors have the same length but point in opposite direction i.e. all their components are with opposite sign (Fig. 8 on the left). They are still correct eigenvectors as can be seen in Fig. 8 on the right, where the absolute values of all eigenvector components are shown. However, sometimes eigenvectors have even different values of their components. Such behavior can be explained by a not unique solution of the eigenvector problem. If some eigenvalues are not distinct, i.e. the solution of the characteristic equation has multiple roots, we say that these eigenvalues are degenerated. Different bases of eigenvectors exist for these degenerate eigenvalues. Therefore, LAPACK and ScaLAPACK can deliver different components for eigenvectors which correspond to degenerate eigenvalues, but they still represent the right eigenvector.

In addition, the correctness of the new subroutine was checked by a comparison of the physical solution for the eigenvectors from LAPACK and ScaLAPACK library. The STARWALL results were in very good agreement within an absolute error of $10^{-13}$.

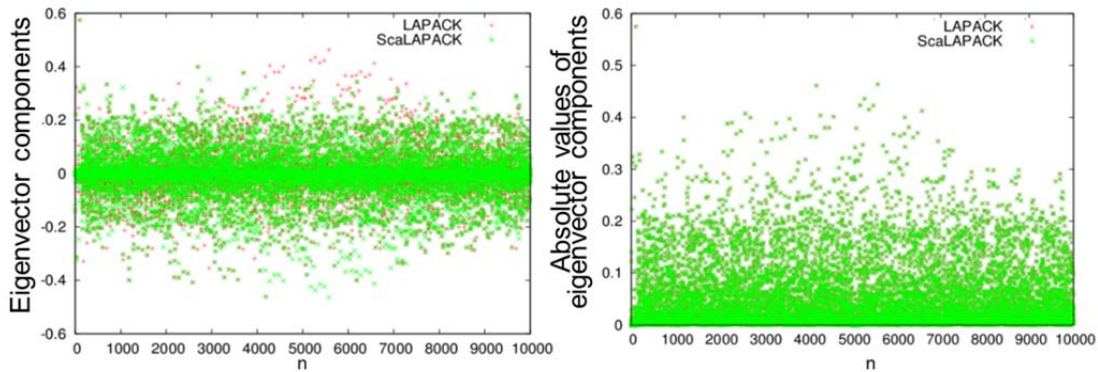

**Fig. 8** Eigenvector components on the left, and their absolute values on the right, from the sequential LAPACK routine *dsygv* (red points) and the parallel ScaLAPACK routine *PDSYGVX* (green points).

The advantage of the ScaLAPACK library in comparison to LAPACK is that it benefits from the IEEE ±∞ arithmetic to accelerate the computations of the eigenvalue solver. Such improvement can be seen in Fig. 9 where the execution time of the ScaLAPACK subroutine *PDSYGVX* obtained from the simulations using one task is compared to the execution time of the LAPACK *dsygv* subroutine for different problem sizes. The ScaLAPACK solver works faster than LAPACK for all problem sizes and gains a factor more than two for large matrices.

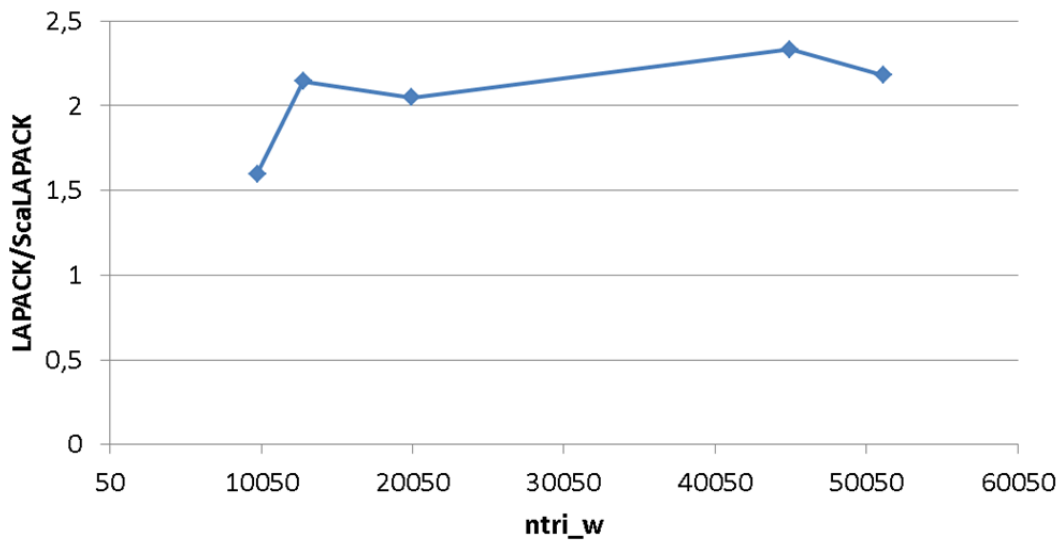

**Fig. 9** Comparison of the eigenvalue solver execution time between ScaLAPACK using one process and the LAPACK library for different problem sizes.

The parallelization efficiency of the *PDSYGVX* subroutine is shown in Fig. 10 on the left for a small problem size (*ntri_w*=10050) and on the right for large matrices (*ntri_w*=51200). For an efficient ScaLAPACK performance the matrix size should be large enough relative to the amount of processes being involved in the simulation [8]. Therefore, the parallelization efficiency is almost saturated with 16 processes for a small problem size with an execution time of only a few seconds. However, when large matrices are used the problem scales almost linearly. An even better performance is expected for a production run in which *ntri_w*=500000.

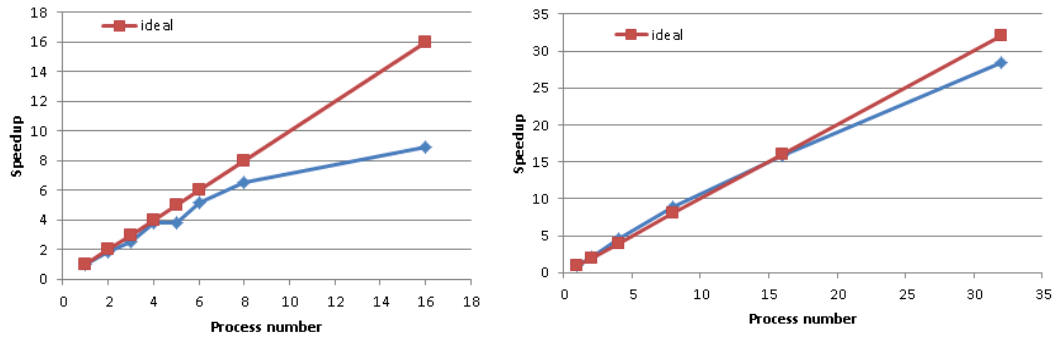

**Fig. 10** *PDSYGVX* parallelization efficiency. On the left, small problem size with *nwu=nwv*=70; on the right, large problem size *nwu=nwv*=160.

### 2.2.2. **Parallelization of the *matrix_ww* subroutine**

The eigenvalue solver described above uses two large matrices (*a_ww(npot_w,npot_w)* and *b_rw(npot_w,npot_w)*) as input parameters. The size of these matrices for a large production run will be (250,000 × 250,000) that is 500 GB for double precision components. Therefore, these matrices have to be distributed over MPI tasks. We started the parallelization with the subroutine *matrix_ww* where the matrix *a_ww* is built.

In this subroutine the matrix *a_ww* is calculated from another matrix, which is named *dima(ntri_w,ntri_w).* The size of this additional matrix is even larger than the size of the matrix *a_ww,* namely (500,000 × 500,000), that is 2 TB for the double precision components. Thus, *dima* matrix must be also distributed over the MPI processes.

The original kernel loop that corresponds to the creation of the matrix *a_ww* is shown in Fig. 11. One can see that the indexes of the matrix *a_ww* and *dima* are not linked. The first one gets its indexes from the additional array *ipot_w* where values range from 1 to *npot_w,* while the *dima* indexes can run from 1 to *ntri_w.*

We tried to find some patterns between the *a_ww* and *dima* matrices such to determine which components of the *dima* matrix will be used for calculating the equally distributed *a_ww* matrix. The *a_ww* matrix was distributed among 16 processors (Fig. 12 left). Each pink rectangle represents the global *a_ww* matrix, and the yellow rectangles depict the sub-matrices assigned to each of the 16 new tasks. The *dima* matrix indexes that were used to calculate the local distributed matrix *a_ww* are shown in Fig. 12 on the right. Now, the pink rectangles stand for the global *dima* matrix, whereas the yellow represent those indexes which are needed to calculate the local part of sub-matrices *a_ww* (yellow rectangles on the left figure). One can see that the *dima* components, which are used to build the distributed part of *a_ww* are not localized and spread across the whole matrix. Hence, it would have been very difficult to efficiently distribute the matrix *dima*.

```
do i =1,ntri_w
    do k =1,3
        j = ipot_w(i,k) + 1
        do i1=1,ntri_w
            do k1=1,3
                j1 = ipot_w(i1,k1) + 1
                temp  = .5*(dxw(i,k)*dxw(i1,k1)                           &
                           +dyw(i,k)*dyw(i1,k1)                           &
                           +dzw(i,k)*dzw(i1,k1))                          &
                           *(dima(i,i1)+dima(i1,i))
                a_ww(j+ncoil,j1+ncoil) = a_ww(j+ncoil,j1+ncoil) + temp
            enddo
        enddo
    enddo
enddo
```

**Fig. 11** Original kernel loop that builds the matrix *a_ww*.

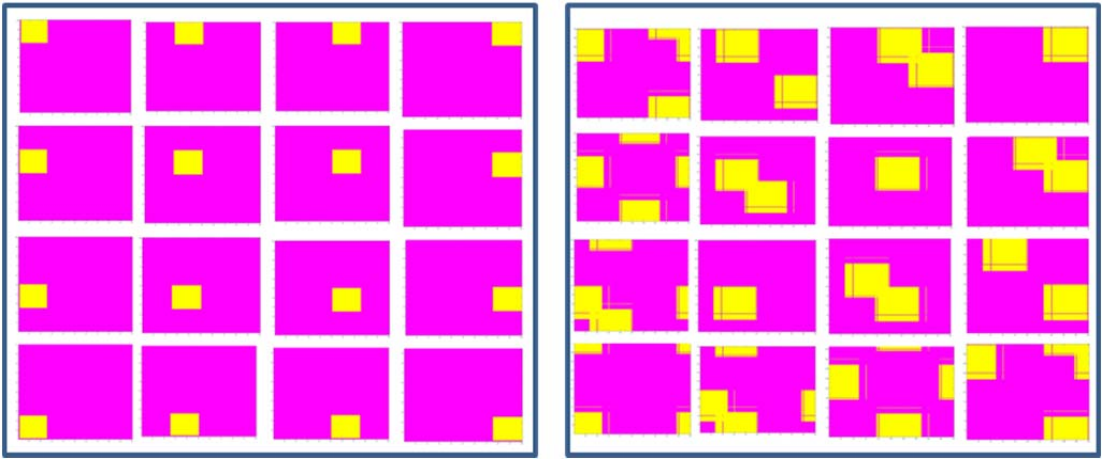

**Fig. 12** Distributed matrix *a_ww* on 16 processors (left) and the corresponding indexes of the matrix *dima* that are used to calculate the local part of *a_ww* (right).

*2.2.2.1. Matrix free "dima" computation*

As the distribution of the matrix *dima* could not be performed efficiently, we decided to rewrite the code in such a way that components of the *dima* matrix will be calculated directly in the place where they should be used.

In the original code version the matrix *dima* was pre-calculated by means of the subroutine *tri_induct,* where three nested loops take place. If this subroutine would be straightforwardly implemented in the kernel loop (Fig. 11), where it has already four nested loops, computational time would be years even on computer clusters. Therefore, we split this subroutine in three parts: *tri_induct_1, tri_induct_2, tri_induct_3.* Two subroutines (*tri_induct_1, tri_induct_2*) are called outside the kernel loop and have no significant effect on the total computational time. Inside the kernel loop only one more nested loop with an index running over seven points was added. A code fragment of the new version of the kernel loop is shown in Fig. 13. One can see that the *dima* matrix is absent there. Instead, there is the function call *tri_induct_3,* where *the* necessary value of *dima* is calculated and stored in the variables *dima_sca* and *dima_sca2*.

The drawback of such a modification is the increase of the computational time. Fig. 14 shows the elapsed time of the kernel loop for different problem sizes using the old version of the code with the matrix *dima* and the new version with the *dima* free format. The computational time increases in about two times for all problem sizes. For a large production run with *ntri_w*=500,000 it was estimated to be around 111 hours on one CPU. The advantage is naturally the possibility to distribute the array and run in parallel.

The next step was to check the parallelization efficiency of the kernel loop. This test is shown in Fig. 15. One can see that a speed-up factor of ~110 can be reached when 256 tasks are involved for the problem size *ntri_w*=12800. Therefore, the computational time of the kernel loop without the *dima* matrix using 256 cores would be about one hour.

```
do i =1,ntri_w
  do i1=1,ntri_w
    do k =1,3
       j = ipot_w(i,k) + 1
       ! If index is inside the local part of distributed matrix a_ww
       if (j>= j_loc_b .AND. j<= j_loc_e ) then
           counter=0
           do k1=1,3
               j1 = ipot_w(i1,k1) + 1
               if (j1>= j1_loc_b .AND. j1<= j1_loc_e ) then
                  dima_sca=0
                  dima_sca2=0
                  if ( counter<1 ) THEN
                      dima_sca=0
                      dima_sca2=0

                      call tri_induct_3(ntri_w,ntri_w,i,i1,xw,yw,zw,dima_sca)
                      call tri_induct_3(ntri_w,ntri_w,i1,i,xw,yw,zw,dima_sca2)

                      dima_sum=dima_sca+dima_sca2
                      counter=counter+1
                  endif

                  temp  = .5*(dxw(i,k)*dxw(i1,k1)               &
                             +dyw(i,k)*dyw(i1,k1)               &
                             +dzw(i,k)*dzw(i1,k1))              &
                             *dima_sum

                  a_ww_loc(j+ncoil,j1+ncoil) = a_ww_loc(j+ncoil,j1+ncoil) + temp

               endif
           enddo
       endif
     enddo
   enddo
enddo
```

**Fig. 13** Matrix *dima* free kernel loop that builds the matrix *a_ww*.

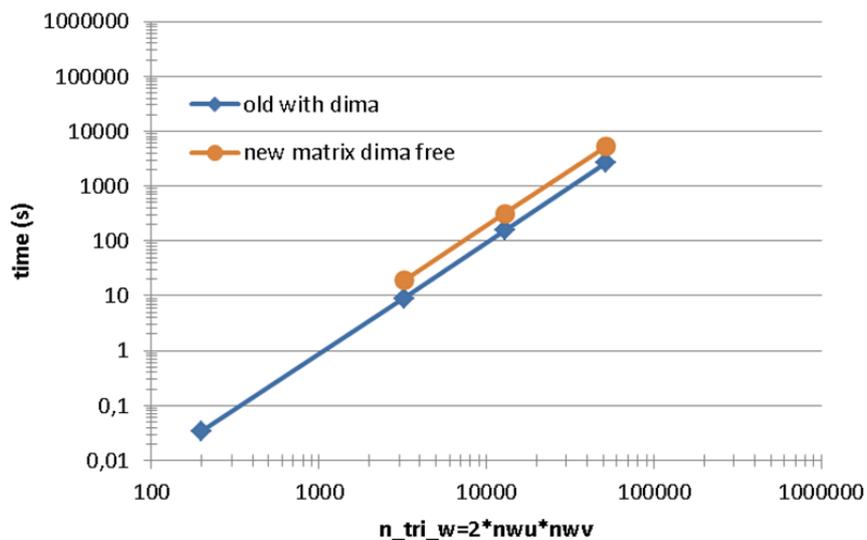

**Fig. 14** Computational time of the kernel loop of the subroutine *matrix_ww* versus the problem size using the old code version (with *dima* matrix) – blue line and modified kernel loop (with *dima* free format) – orange line.

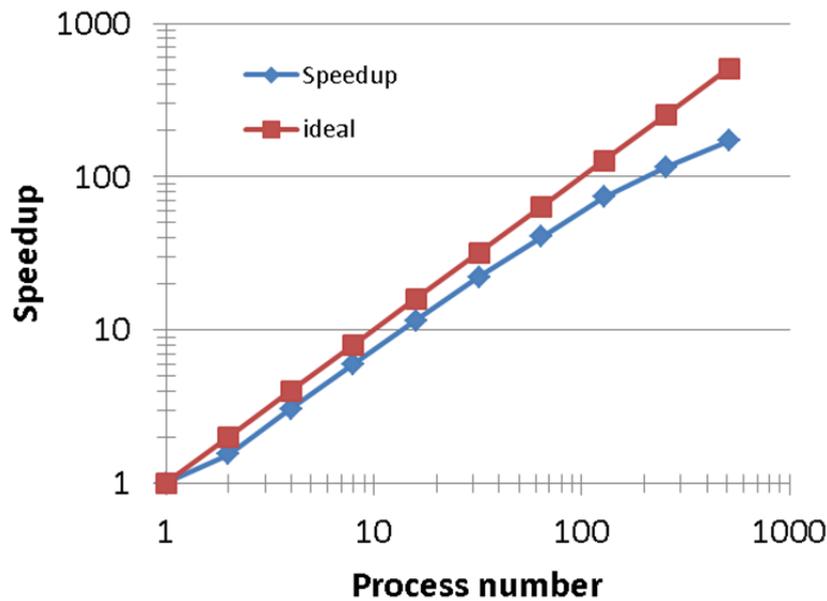

**Fig. 15** Speed-up of the kernel loop versus number of MPI tasks. The problem size is *ntri_w*=12800

### 2.2.2.2. Matrix free "dima" computation with ScaLAPACK indexing

In order to use the distributed matrices as input parameters for ScaLAPACK subroutines they must be transformed to a special format using the so-called Block-Cyclic distribution scheme, which should speed-up the calculation [8]. For example, if we consider the global matrix with a size of 9×9, which is mapped onto a 2×3 process grid (six tasks) and with a blocking factor of two, the decomposition which is shown in Fig. 16 has to be done. On can see that in this format different processes have different local matrix sizes, from 5×4 for process (0,0) to 4×2 for process (1,2). Moreover, the mapped indexes in the local distributed matrix are not sequential. For instance, in the process (0,0) the first row includes the following elements of the global matrix: $a_{11}$, $a_{12}$, $a_{17}$, $a_{18}$.

|   |   | 0 | | | | 1 | | | 2 | |
|---|---|---|---|---|---|---|---|---|---|---|
| 0 |   | $a_{11}$ | $a_{12}$ | $a_{17}$ | $a_{18}$ | $a_{13}$ | $a_{14}$ | $a_{19}$ | $a_{15}$ | $a_{16}$ |
|   |   | $a_{21}$ | $a_{22}$ | $a_{27}$ | $a_{28}$ | $a_{23}$ | $a_{24}$ | $a_{29}$ | $a_{25}$ | $a_{26}$ |
|   |   | $a_{51}$ | $a_{52}$ | $a_{57}$ | $a_{58}$ | $a_{53}$ | $a_{54}$ | $a_{59}$ | $a_{55}$ | $a_{56}$ |
|   |   | $a_{61}$ | $a_{62}$ | $a_{67}$ | $a_{68}$ | $a_{63}$ | $a_{64}$ | $a_{69}$ | $a_{65}$ | $a_{66}$ |
|   |   | $a_{91}$ | $a_{92}$ | $a_{97}$ | $a_{98}$ | $a_{93}$ | $a_{94}$ | $a_{99}$ | $a_{95}$ | $a_{96}$ |
| 1 |   | $a_{31}$ | $a_{32}$ | $a_{37}$ | $a_{38}$ | $a_{33}$ | $a_{34}$ | $a_{39}$ | $a_{35}$ | $a_{36}$ |
|   |   | $a_{41}$ | $a_{42}$ | $a_{47}$ | $a_{48}$ | $a_{43}$ | $a_{44}$ | $a_{49}$ | $a_{45}$ | $a_{46}$ |
|   |   | $a_{71}$ | $a_{72}$ | $a_{77}$ | $a_{78}$ | $a_{73}$ | $a_{74}$ | $a_{79}$ | $a_{75}$ | $a_{76}$ |
|   |   | $a_{81}$ | $a_{82}$ | $a_{87}$ | $a_{88}$ | $a_{83}$ | $a_{84}$ | $a_{89}$ | $a_{85}$ | $a_{86}$ |

**Fig. 16** Example of the Block-Cycling matrix distribution of size 9×9 into 2×2 blocks mapped onto a 2×3 process grid.

Hence, the Block-Cyclic distribution scheme described above has to be implemented in the subroutine *matrix_ww* in order to bring the local distributed matrix *a_ww* to a format compatible with the ScaLAPACK subroutines. Such index mapping was developed and implemented in two subroutines: *ScaLAPACK_mapping_i*, *ScaLAPACK_mapping_j* and then inserted in the kernel loop. Such index distribution causes bad scalability of the kernel loop when using the same structure shown in Fig. 13. Therefore, this kernel loop was rewritten one more time to ensure good scalability with the ScaLAPACK mapping scheme (Fig. 17). Using 512 cores with the new version a speed-up factor of 218 could be reached. The wall clock time was estimated for a large production run with *ntri_w*=500,000 to be about 4 hours.

```
do i =1,ntri_w
   do i1=1,ntri_w
      do k =1,3
         j = ipot_w(i,k) + 1
         call  ScaLAPACK_mapping_i(j,i_loc,inside_i)
         if (inside_i == .true.) then

            do k1=1,3
               j1 = ipot_w(i1,k1) + 1
               call ScaLAPACK_mapping_j(j1,j_loc,inside_j)
               if (inside_j == .true.) then

                  dima_sca=0
                  dima_sca2=0
                  call tri_induct_3(ntri_w,ntri_w,i,i1,xw,yw,zw,dima_sca)
                  call tri_induct_3(ntri_w,ntri_w,i1,i,xw,yw,zw,dima_sca2)

                  dima_sum=dima_sca+dima_sca2

                  temp   = .5*(dxw(i,k)*dxw(i1,k1)              &
                          +dyw(i,k)*dyw(i1,k1)                  &
                          +dzw(i,k)*dzw(i1,k1))                 &
                          *dima_sum

                  a_ww_loc(i_loc,j_loc) = a_ww_loc(i_loc,j_loc) + temp

               endif
            enddo
         endif
      enddo
   enddo
enddo
```

**Fig. 17** ScaLAPACK index mapping *dima* free kernel loop that builds the matrix *a_ww*.

### 2.2.3. **Parallelization of the *matrix_pp* subroutine**

The next subroutine chosen for parallelization was *matrix_pp*. It produces the intermediate matrix (*a_pp*) that will be used to calculate the input matrix for the eigenvalue solver. This subroutine is similar to the *matrix_ww* described above. The main difference lies in the construction of the *dima* matrix. It uses two additional matrices *dist1* and *dist2* in order to calculate its components. The size of the *dima* and the resulting matrix *a_pp* is also different from the previous subroutine, because it corresponds to the number of triangles within the plasma that should be discretized by *ntri_p*=200000 for a large production run. On one side, we got more complexity in the kernel loop, on the other side, the loop is smaller in comparison to the kernel *matrix_ww*.

The additional subroutine (*get_index_dima*) was developed in order to determine which indexes of the matrix *dima* are used for computing the matrix *a_pp* components. The kernel loop of this subroutine is shown in Fig. 18.

The scalability of this kernel loop, depicted in Fig. 18, is shown in Fig. 19. A speed-up factor of 220 can be achieved when 512 cores are involved in the computation for the problem size *ntri_p*=46080. For a large production run the wall clock time (with 512 cores and *ntri_p*=200,000) reduces to about 2 hours.

```
do i =1,ntri_p
   do i1=1,ntri_p
      do k =1,3
      j = ipot_p(i,k) + 1
       call  ScaLAPACK_mapping_i(j,i_loc,inside_i)
       if (inside_i == .true.) then

          do k1=1,3
          j1 = ipot_p(i1,k1) + 1
          call ScaLAPACK_mapping_j(j1,j_loc,inside_j)
          if (inside_j == .true.) then

             call get_index_dima(i,i1,ku,ku2)

             dima_sca=0
             dima_sca2=0
             call tri_induct_3(ntri_p,ntri_p,ku,ku2,xp,yp,zp,dima_sca)
             call tri_induct_3(ntri_p,ntri_p,ku2,ku,xp,yp,zp,dima_sca2)
             dima_sca3=.5*(dima_sca+dima_sca2)

             temp  = (dxp(i,k)*dxp(i1,k1)                              &
                   +  dyp(i,k)*dyp(i1,k1)                              &
                   +  dzp(i,k)*dzp(i1,k1)) *dima_sca3

             a_pp_loc(i_loc,j_loc) = a_pp_loc(i_loc,j_loc) + temp

          endif
          enddo
       endif
       enddo
    enddo
 enddo
enddo
```

**Fig. 18** ScaLAPACK index mapping *dima* free kernel loop that builds the matrix *a_pp* in subroutine *matrix_pp*.

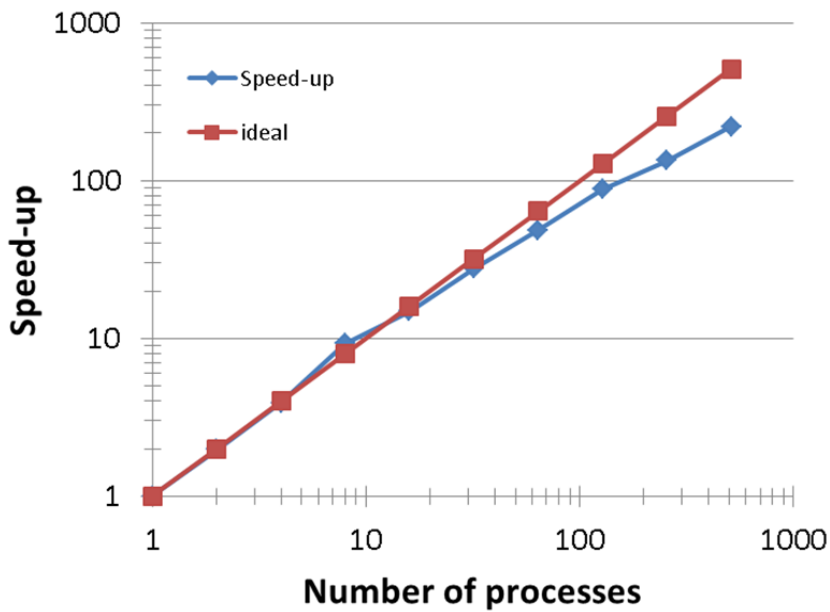

**Fig. 19** Speed-up of the kernel loop in the *matrix_pp* subroutine versus number of MPI tasks. The problem size is *ntri_p*=46080.

### 2.2.4. **Parallelization of the *matrix_wp* subroutine**

The *matrix_wp* subroutine is similar to the previously parallelized subroutines *matrix_ww* and *matrix_pp* described above. The main difference lies in the presence of two large matrices, *dima* and *dimb,* that have to be eliminated from the code in order to save a significant amount of memory. Therefore, the components of these two matrices have to be calculated directly in place rather than stored in memory.

Additionally, the *a_wp* matrix size (*npot_w*, *npot_p*) and the indexes of the kernel loop (*ntri_w*, *ntri_p*) are also different from the previous subroutines.

The subroutine was successfully parallelized providing identical results as the original version within an absolute difference of $10^{-10}$. The scalability of the subroutine is shown in Fig. 20. A speed-up factor of 148 can be achieved when 256 cores are involved in the computation. The subroutine was tested for a large production run with *ntri_p*=$2*10^5$ and *ntri_w*=$5*10^5$. The execution time with 128 tasks was about 3.5 hours.

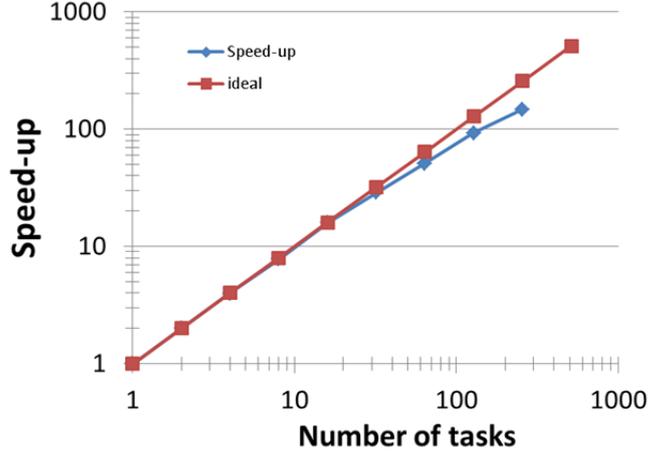

**Fig. 20** Speed-up of the *matrix_wp* subroutine versus number of MPI tasks.

### 2.2.5. Parallelization of the *matrix_rw* subroutine

The parallelization of the *matrix_rw* subroutine was relatively straightforward in comparison to the previous *matrix_wp* subroutine since it does not involve the large matrices *dima* and *dimb*. The only problem was to bring the local matrix *a_rw* to the ScaLAPACK matrix structure described earlier. The subroutine was successfully parallelized providing accurate results within difference of ~$10^{-10}$. The subroutine was tested for a large production run with *ntri_p*=$2*10^5$ and *ntri_w*=$5*10^5$. The execution time using 256 tasks was in the range of a few minutes.

### 2.2.6. Parallelization of the *matrix_pe* subroutine

The *matrix_pe* subroutine has a different kernel loop structure compared to all previously parallelized subroutines. It is independent of the *dima* and *dimb* matrices and the indexes of the kernel loop run from 1 to the number of harmonics (*n_harm*) and to the number of boundary elements (*N_bnd*). The subroutine was parallelized with high accuracy (absolute difference of ~$10^{-10}$) and the output matrix (*a_pwe*) was re-ordered to be compatible with the ScaLAPACK matrix structure. Because of much smaller values of *n_harm* and *N_bnd* than *ntri_p* and *ntri_w* the execution time for this subroutine is small (few minutes) for a production run.

### 2.2.7. Parallelization of the *matrix_ep* and *matrix_ew* subroutines

The subroutines *matrix_ep* and *matrix_ew* have a similar structure with differences only in the size of the main arrays (*a_ep* and *a_ew*). *a_ep* has the size of the potential points for the plasma (*npot_p*) and *a_ew* of the potential points for the wall (*npot_w*). All other loops and components are identical.

In the main body of these subroutines three additional supplying subroutines are called. They are *bfield_par, bfield_c* and *real_space2bezier*. Moreover, inside the subroutine *real_space2bezier* two LAPACK functions are executed (*dpotrf* and *dpotrs)*. The former computes the lower-upper (LU) factorization of a tridiagonal matrix, while the latter solves a system of linear equations with a Cholesky factored symmetric positive definite matrix. Fortunately, these functions use as input parameters the matrices *aa* and *t* with dimensions (*n_dof_bnd*, *n_dof_bnd*). As the variable *n_dof_bnd* is about 400 for a production run, the double precision arrays (*aa* and *t*) will not represent more than 1.5 MB. Therefore, we left these LAPACK functions untouched i.e. in the sequential version.

After the parallelization of the subroutines *matrix_ep* and *matrix_ew,* including the inner supplying subroutines, the total computational time was measured for a production run with *ntri_w*=500000. Using 256 tasks the wall clock time for the *matrix_ep* was 51 s, while 15 s was necessary for computing the *matrix_ew* subroutine.

### 2.2.8. Parallel matrix transpose

One part of the STARWALL solver recalculates the entries of the matrix *a_pwe* by using values from the transposed matrix *a_wp.* In order to improve the code performance this subroutine was replaced by the ScaLAPACK library function *PDTRAN* that can be adapted for a matrix transpose. The wallclock time does not exceed a few seconds for the production run.

### 2.2.9. Parallel LU factorization with linear system solver

Two LAPACK functions named *dpotrf* and *dpotrs* are executed after the *a_pwe* matrix transpose. The first function computes the lower-upper (LU) factorization of a tridiagonal matrix *a_pp*, while the second solves a system of linear equations with a Cholesky factored symmetric positive definite matrix. Both functions were replaced with their parallel counterpart from the ScaLAPACK library and grouped in the subroutine *cholesky_solver.* The subroutine provides the correct result within an absolute error of $10^{-10}$ in comparison with the sequential LAPACK version.

### 2.2.10. Parallelization of building matrix *a_ee*

The sequential version of the code for building the matrix *a_ee* is shown in Fig. 21. As one can see this matrix is formed by the multiplication of the matrices *a_ep* and *a_pwe* using only a small part of the elements of the matrix *a_pwe*. This loop was replaced by the ScaLAPACK subroutine named *PDGEMM* that computes the matrix-matrix product. However, before the execution of this subroutine the distributed matrix *a_pwe* was rewritten to be used in the ScaLAPACK *PDGEMM* subroutine. Finally, the parallel version of the building matrix *a_ee* was tested and it provided correct results compared to the sequential version.

```
do i=1,nd_bez
   do k=1,nd_bez
      do j=1,npot_p
         a_ee(i,k) = a_ee(i,k)+a_ep(i,j)*a_pwe(j,k+nd_w)
      enddo
   enddo
enddo
```

**Fig. 21** Sequential version of building the matrix *a_ee.*

### 2.2.11. Parallelization of building matrices *a_ew* and *a_we*

Fig. 22 shows the sequential version of the building of the matrices *a_ew* and *a_we.* The structure of these loops is similar to the one described in the previous section with different sizes and indices. However, both loops can be replaced by the ScaLAPACK subroutine for the matrix-matrix product (*PDGEMM*) as it was done for building the matrix *a_ee*. Two new subroutines named *a_ew_computing* and *a_we_computing* were created, which include the parallel building of the distributed matrices *a_ew* and *a_we*, respectively.

```fortran
do i=1,nd_bez
  do k=1,nd_w
    do j=1,npot_p
      a_ew(i,k) = a_ew(i,k) - a_ep(i,j)*a_pwe(j,k)
    enddo
  enddo
enddo

do i=1,nd_w
  do k= 1,nd_bez
    do j=1,npot_p
      a_we(i,k) = a_we(i,k) +a_wp(i,j)*a_pwe(j,k+nd_w)
    enddo
  enddo
enddo
```

**Fig. 22** Sequential version of building the matrices *a_ew* and *a_we*.

### 2.2.12. Parallelization of the LAPACK *dgemm* subroutine

The last call of the STARWALL *solver* subroutine is the LAPACK *dgemm* subroutine for the multiplication of the matrices *a_wp* and *a_pwe*. This subroutine was replaced by its parallel counterpart from the ScaLAPACK library namely *PDGEMM*. The same subroutine was used to build the matrices *a_we*, *a_ew* and *a_ee.* Therefore, its implementation was relatively easy and required only a few additional ScaLAPACK descriptors. The whole computation was encapsulated in the subroutine named *matrix_multiplication.*

### 2.2.13. Parallelization of *resistive_wall_response* subroutine

The *resistive_wall_response* subroutine follows after the *solver* subroutine described above. There are three main parts of this subroutine: (i) eigenvalue solver, (ii) preparation of output matrices and (iii) printing of final results. The eigenvalue solver has been parallelized in the very beginning of this project described in section 2.2.1.

After solving for the eigenvalues the output matrices *a_ye*, *a_ey* and *d_ee* are computed. The sequential version of the calculation of these matrices is presented in Fig. 23. As we can see the matrices *a_ey* and *d_ee* are computed by the matrix-matrix multiplication scheme, while in order to calculate the matrix *a_ye* the transpose of the matrix *s_ww* is required. All loops were successfully parallelized and copied in three subroutines named *a_ey_computing*, *a_ye_computing* and *d_ee_computing.*

The last part of the *resistive_wall_response* subroutine is printing the computed matrices to the different output files. All matrices that were calculated in the parallel version of the STARWALL code are distributed over the number of MPI tasks using the ScaLAPACK block-cycling distribution scheme. Thus, the output subroutine should match with the reading subroutine in the JOREK code that is not implemented yet. Therefore, we did not modify the printing part of the code and postpone it until the reading part in JOREK will be implemented in order to know the necessary output format.

```fortran
a_ye =0.
do i=1,n_w
   do k=1,nd_bez
      do j=1,n_w
         a_ye(i,k) = a_ye(i,k) + S_ww(j,i)*a_we(j,k)
      enddo
   enddo
enddo

do i=1,n_w
   do k=1,nd_bez
      a_ye(i,k) = a_ye(i,k)/gamma(i)
   enddo
enddo

a_ey = 0.
do i=1,nd_bez
   do k=1,n_w
      do j=1,n_w
         a_ey(i,k) = a_ey(i,k) + a_ew(i,j)*S_ww(j,k)
      enddo
   enddo
enddo

d_ee = 0.
do i=1,nd_bez
   do k=1,nd_bez
      do j=1,n_w
         d_ee(i,k)= d_ee(i,k) + a_ey(i,j)*a_ye(j,k)
      enddo
   enddo
enddo
```

**Fig. 23** Sequential version of computing the final matrices *a_ye*, *a_ey* and *d_ee*.

### 2.2.14. Parallelization of matrix *s_ww* inversion

The last computing subroutine of the STARWALL code, before printing out the final results, performs the inversion of the eigenvectors matrix (*s_ww*). Two LAPACK subroutines are used for this purpose. They were replaced by their parallel counterpart from the ScaLAPACK library. First, the subroutine named *PDGETRF* calculates the LU factorization of a general matrix using partial pivoting. Second, *PDGETRI* computes the inverse of a matrix using LU factorization from the previous step. Both subroutines were grouped in the subroutine named *computing_s_ww_inverse*. The computational time of this subroutine was measured to be of ~2805 s for a production run (*ntri_p*=2*$10^5$ and *ntri_w*=5*$10^5$).

### 2.2.15. Parallelization of input subroutines

Three input subroutines were also parallelized: *control_boundary* that reads the JOREK control boundary data; *tri_contr_surf* that is used to generate the control surface triangles and *surface_wall* that performs the discretization of the wall. These subroutines were parallelized in such a way that only one master task reads the data from the input files and broadcasts it to the tasks involved in the computation. An additional subroutine named *control_array_distribution* was inserted after the reading part. This subroutine controls and checks the distribution of the matrices among the MPI tasks.

## 2.3. *Parallel performance test*

After the whole code was parallelized and tested for the correctness of the output results we did a comparison of the code performance with respect to the original version. The maximum possible problem size for the original code version which fits into memory is the following: *ntri_p*=48000, *ntri_w*=65000, *nharm*=11 (57 GB memory consumption). The wallclock time for such a simulation using 16 OpenMP processes is ~4 hours. We performed a simulation with identical parameters but with the new (MPI parallel) code version. In spite of the larger complexity of the solver due to the new version of the matrix building subroutines, which avoids the storing of the largest matrices in the code named *dima* and *dimb,* the total computational time (excluding the output) on one computing node and 16 MPI tasks is about the same as it is in the OpenMP version of ~4.2 hours consuming 41 GB of the memory. However, the computational time is reduced to about 40 minutes when using eight compute nodes and 128 MPI tasks. Nevertheless, for the small problem sizes which fit in the memory of one node, the OpenMP version is faster than the parallel one with 16 MPI tasks.

Next step was to test the code performance for a typical production run with the following parameters: *ntri_p*=202.240, *ntri_w*=500.000, *nharm*=11. Fig. 24 shows the execution time of some subroutines from the parallel version of the STARWALL code. For this test 2048 MPI tasks were used distributed among 128 compute nodes on HELIOS. The execution time from all subroutines shown in Fig. 24 represents 99% of the total computational time that is about 11 hours. One can see that four subroutines (*matrix_pp*, *matrix_wp*, *matrix_ww* and the eigenvalue solver – *simil_trafo*), described in details above, consume most of the computational time.

| Name | Computation time (s) |
|---|---|
| matrix_pp | 2774 |
| matrix_wp | 7591 |
| matrix_ww | 13206 |
| matrix_rw | 0,27 |
| matrix_pe | 0,007 |
| matrix_ep | 189 |
| matrix_ew | 146 |
| cholesky_solver | 345 |
| a_pwe_s_computing | 1591 |
| a_ee_computing | 3,9 |
| a_ew_computing | 38 |
| a_we_computing | 34 |
| matrix_multiplication (dgemm) | 382 |
| simil_trafo (Eigenvalue solver) | 11820 |
| a_ye_computing | 49 |
| a_ey_computing | 63 |
| d_ee_computing | 75 |

**Fig.** 24 The wall clock time of some subroutines from the parallel version of the STARWALL code for a production run with the following parameters: *ntri_p*=202.240, *ntri_w*=500.000, *nharm*=11. The subroutines are listed in their execution order.

We gradually increased the problem size and determined the maximum possible run within 128 nodes with the following parameters: *ntri_p*=202.240, *ntri_w*=551.250, *nharm*=11 and a wallclock time about 13 hours.

### 2.3.1. Parametric scan of the ScaLAPACK blocking factor

It was mentioned above that the ScaLAPACK library requires a special matrix distribution format (Block-Cyclic). The blocking size of such a format is defined by the user, and it has a strong impact on the code performance. Fig. 25 shows the wall clock time for a production run ($ntri\_p$=202.240, $ntri\_w$=500.000) of five subroutines, that consume more than 95% of the total computational time, versus different sizes of the ScaLAPACK blocking factor (from $NB$=2 to $NB$=256). Among these subroutines are two from the ScaLAPACK library (matrix multiplication – *DGEMM* and the eigenvalue solver – *PDSYGVX*) and three for the building matrices (*matrix_pp*, *matrix_wp*, *matrix_ww*). One can see that the execution time of the ScaLAPACK subroutines decreases using a higher blocking factor. For small blocking factors ($NB$=2 or $NB$=4) the execution time of the eigenvalue solver is too large (>15 hours) for the program to finish within 24 hours. Therefore, these points are not depicted in Fig. 25. A significant reduction of the computational time is visible up to $NB$=64. After that the execution time decreases but only by a few percent when it reaches $NB$=128. With $NB$=256 the execution time of these subroutines begins to increase. The computational time of the matrix building subroutines fluctuates for all blocking factors. The total computational time (orange line) shows that the best performance for such a problem size is ~11 hours with $NB$=64. This is in agreement with the ScaLAPACK documentation where developers propose for the best performance to use the following blocking factors $NB$=32, 64 or 128 [9]. However, for a different problem size the best performance could be with a different blocking factor. Therefore, the STARWALL input file was extended including now the blocking factor as an input parameter.

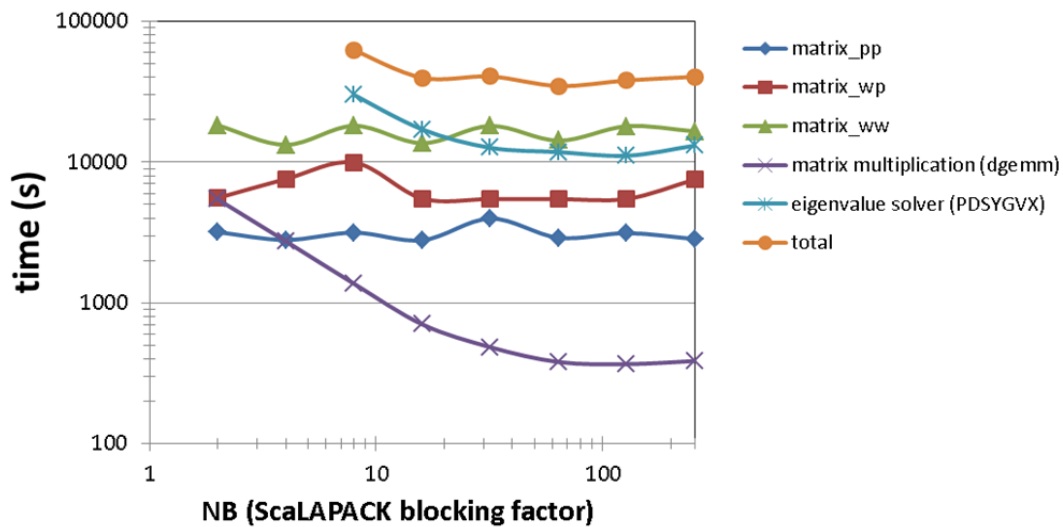

**Fig. 25** The wall clock time versus the ScaLAPACK blocking factor for a production run with the following parameters: $ntri\_p$=202.240, $ntri\_w$=500.000, $nharm$=11.

### 2.3.2. Scalability test

We tested also how the total computational time scales according to the number of MPI tasks involved in the calculation. We decreased the problem size to be able to run it on a smaller number of compute nodes. Fig. 25 shows the wall clock time for a production run ($ntri\_p$=202.240, $ntri\_w$=460.800) of the four subroutines, that consume most of the total computational time, versus the number of MPI tasks. For such a problem size the whole code can be executed on 64 nodes (1024 MPI tasks). With a smaller number of nodes only a part of the code is performing due to the memory limit. One can see that the wall clock time decreases for all subroutines up to 128 nodes (2048 MPI tasks). Up to 4096 tasks the execution time of the ScaLAPACK eigenvalue solver continues to shrink. However, the computational time of the three matrix building subroutines starts to grow above 2048 tasks. We detected that the optimal code performance could be reached by using 128 compute nodes with a ScaLAPACK blocking factor of 64 for the production run described above. For a larger or smaller problem size the scaling could be different.

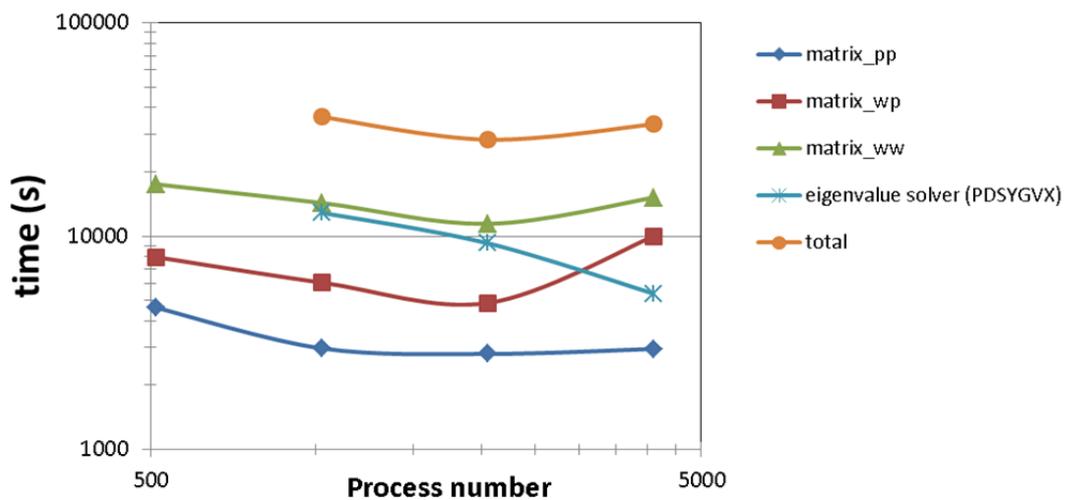

**Fig.** 26 Scaling of the most time consuming subroutines in the STARWALL code.

The scalability of the whole program execution including the output was tested also for a moderate problem size with *ntri_p*=48000, *ntri_w*=39200, *nharm*=11 (Fig. 27). A speed-up factor of nine was achieved with 256 MPI tasks in comparison to 16 MPI tasks. On a node, the original version is faster than the parallel one due to the much more complex algorithm used for the matrix building subroutines that avoids to store the largest matrix in the code named *dima* and *dimb*. However, with two nodes the total wall clock time becomes smaller than in the original version and the speed-up factor of six can be achieved with 256 MPI tasks in comparison to the original version.

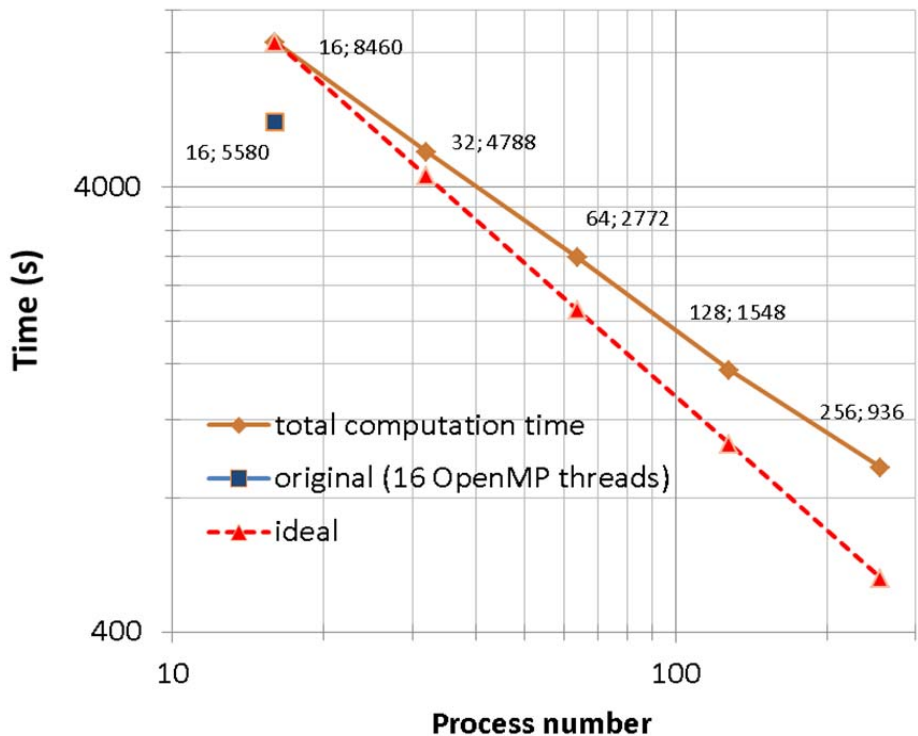

**Fig. 27** Scaling of the total wallclock time in the STARWALL code for a small problem size with *ntri_p*=48000, *ntri_w*=39200, *nharm*=11. The numbers next to the points are the task number and computation time.

### 2.3.3. Temporary output

In the future a consistent format of the output of the STARWALL code and the input of the JOREK code has to be chosen. After that both subroutines must be parallelized. At the moment we use the same output format in the parallel version as in the sequential one. This gives a limitation for the problem size resulting from the output matrix size of no more than 3.5 GB due to the memory capacity of the node of 64 GB and assuming that we run 16 MPI tasks per node where each task should allocate such output matrix.

## 2.4. *Parallelization of the code version for magnetic coils*

The standard code version does not include a calculation over the external magnetic coils. However, in the future, this feature of the code must be usable for production runs with a high number of finite element triangles. Therefore, it was decided to parallelize the subroutines that deal with the magnetic coils. Among them are one reading (*read_coil_data*) and five matrix building subroutines (*matrix_cc*, *matrix_cp, matrix_wc*, *matrix_rc, matrix_ec*). All these subroutines have been successfully parallelized providing identical results in comparison with the original code version. Due to the project time limit the performance of these subroutines was not measured. However, it is expected that including these additional subroutines will not increase the wallclock time for a production run by more than 10–20 %. This is due to the relative small matrix sizes being involved in the external coils calculation in comparison to the matrices that were parallelized before.

# 3. MPI parallelization of the magnetohydrodynamics code JOREK

The JORSTAR2 project is a continuation of the JORSTAR project described above and dedicated to the implementation of parallel I/O in the STARWALL output and JOREK input subroutines. The large STARWALL matrices are distributed over MPI tasks to reduce memory consumption and to allow for running larger simulations in terms of the JOREK computational grid and the number of triangles used in STARWALL to discretize wall structures. A sequential part of JOREK in which the input matrices from the STARWALL code are used has to be parallelized as well.

## 3.1. *Goal of the project*

Thanks to the JORSTAR project it is now possible to resolve the realistic wall structure with a large number of finite element triangles in the STARWALL code. However, the output subroutine is still sequential. This project concentrates on the MPI parallelization of the sequential I/O part in both the JOREK and STARWALL codes and adapting the JOREK code for using the STARWALL response matrices now distributed over MPI tasks.

## 3.2. *Parallelization of the STARWALL output subroutine*

Before starting the implementation of the parallel I/O modules in both STARWALL and JOREK a variety of libraries and subroutines were analyzed in order to find the best candidate.

Most linear algebra subroutines were parallelized in the previous JORSTAR project by means of the parallel ScaLAPACK library [10]. This library requires the so-called block-cycling matrix distribution format described in detail in Sec. 2.2.2. Before they can be used in an output procedure, all these matrices have to be converted to a standard contiguous format. We first investigated if the ScaLAPACK library has a suitable subroutine for parallel I/O with a direct conversion to this standard format. Only one subroutine named *PDLAPRNT* was found. This subroutine collects all distributed local matrices, converts them from the block-cycling distribution format to the contiguous one and writes a global matrix. It could have been a straightforward solution for our problem as everything in the STARWALL code is already prepared for the ScaLAPACK library. However, after applying and testing this subroutine we realized that it only works for small problem sizes. The subroutine is written in such a way that only one MPI task locally collects all distributed matrices and afterwards writes them to a file. This makes the output very slow and restricts the maximum global matrix size to the memory capacity of one computing node or less (<180 GB on the Skylake partition of the Marconi supercomputer). Therefore, this subroutine does not meet our requirements as some global matrices of a STARWALL production run can have sizes of about 500 GB.

The next step was to test the ROMIO library which is an implementation of the MPI 3.0 standard. This library includes many different MPI I/O subroutines which were tested for our project. We started with *MPI_File_seek, MPI_File_write* and *MPI_File_write_at*. The first subroutine seeks to the writing position, while the second subroutine performs the writing itself. The last subroutine is a combination of the first two. These subroutines were working fine and provided the correct output. However, they are not collective routines which makes the output very slow for large problem sizes (a couple of days for a matrix larger than 500 GB). Moreover, they require an additional calculation for transforming the block-cycling distribution to the contiguous format. Therefore, these subroutines are not suitable for our project.

Next, we tested the collective subroutine *MPI_File_write_at_all,* which has the same functionality as the previously described subroutines with the difference that all MPI tasks write simultaneously to a file. Again this subroutine works fine and much faster than the previous ones but it also has one restriction. Each MPI task has to call it the same number of times as it is a collective subroutine but sometimes the sizes of the local distributed matrices are not identical for each MPI task. Therefore, in such cases the subroutine gets stuck and the program deadlocks.

Finally, a solution was found by using the subroutines *MPI_Type_create_darray* and *MPI_File_set_view*. The former creates a description of any complex data structure, for example block-cycling distributed submatrices, while the latter defines an independent file view for each MPI task. Therefore, each MPI task can write its own specific data structure concurrently to the same file by means of a single call to the *MPI_File_write_at_all* subroutine described before. This method works very fast for any problem size and delivers correct results. Similar routines can be used in JOREK to provide a different distribution of matrices over MPI tasks already when reading them. This is described in the following section.

However, we decided to continue to investigate further possible candidates for our problem and tested the parallel HDF5 library as it was already successfully used in some parts of the JOREK code. We were able to achieve correct and fast performance for equally distributed matrices (each submatrix has the same size). However, we did not find a possibility by means of the HDF5 library to in parallel write not equally block-cycling distributed matrices. Therefore, we kept the solution described in the previous paragraph.

### 3.2.1. STARWALL parallel I/O performance test

After implementing the solution for each output matrix in STARWALL, performance measurements were conducted. Different problem sizes as well as different numbers of computing nodes involved in the writing process were tested. All tests were done on the Broadwell partition of the Marconi supercomputer, which offers 36 cores per node. The wall clock time for the complete output was about 83 seconds for a moderate problem size with *ntri_w*=120,000 finite element triangles in the wall using two computing nodes. The output file size for this case was about 110 GB. Only ~100 seconds were needed for a production run output with *ntri_w*=500,000 and *ntri_p*=202,000 using 64 computing nodes creating a file with a size of about 1 TB. A complete STARWALL production run, including all computations and output procedures, with *ntri_w*=500,000 and *ntri_p*=200,000 using 64 computing nodes takes about nine hours. This is much less time than is required by the project coordinator (<24 hours). At this step all work concerning the STARWALL code is finished.

### 3.3. *Parallelization of the JOREK input subroutine*

The same solution that was used for the STARWALL output was applied to the JOREK input subroutine including only small modifications which are described here. The global matrices should not be read from the input file (i.e. the STARWALL output file) in the block-cycling distribution format but in an ordinary way using a row– or a column–wise distribution. Therefore, instead of the *MPI_Type_create_darray* subroutine, the *MPI_Type_create_subarray* subroutine was used. The data structure of the matrices was also modified. Instead of using a simple allocatable array, we introduced a data type that includes local allocatable submatrices, the starting and ending indices of the global matrix and the type of distribution (column–or row–wise).

Results obtained from the new parallel reading subroutine were compared with the old sequential version. They were identical up to a relative error of ~$10^{-13}$. Afterwards the execution time of the whole reading procedure for the production run (*ntri_w*=500,000, *ntri_p*=200,000) was measured. Using 16 computing nodes and 36 MPI tasks per node the wall clock time was less than one minute.

### 3.4. *Restrictions of the Intel MPI 3.0 library*

One important limitation in the subroutines *MPI_File_write_at_all* and *MPI_File_read_at_all* was found during the development of the parallel MPI I/O. The amount of elements to be read/written from/to a file by each MPI task is an input parameter for both subroutines. According to the Intel MPI documentation [11] this variable is a four byte integer that can have a maximum value of 2147483647. This corresponds to approximately two GB of data. For double precision (eight bytes per value) arrays, which are used in our codes, the maximum size of data that can be read/written by each MPI task is limited to 2147483647*8 bytes ≈ 16 GB. This means that if an array size is larger than *number_of_MPI_tasks*\*16 GB the code will fail. For a production run, the largest output matrix uses about 500 GB. Therefore, we need a minimum of 32 MPI tasks in order to overcome this limitation and to perform the correct reading/writing procedure. The next MPI standard, 4.0, should correct this limitation by changing the data type of the count variable. However, it was decided to introduce modifications in the reading subroutine in order to avoid this limitation. If the local matrix size is larger than 16 GB, the reading procedure will be performed in several steps, reading a data chunk that is less than 16 GB on each step.

As described above, and according to the MPI 3.0 standard, it should be possible to read 16 GB of data per MPI task in one operation for a double precision array. However, in the Intel MPI implementation of these subroutines, the variable *count* is multiplied by the type of the read array (eight for double precision) and the result of this operation is stored in a four byte integer variable. Therefore, if we try to read the maximum possible amount of elements (2147483647) for a double precision data type the code crashes with a segmentation fault. As long as this bug is present in the Intel MPI library we need to restrict our reading chunk size to less than two GB. This bug was reported to the Intel support team [12] and should be fixed in the next version of the library.

Additional modifications were made which ensure that each MPI task can read a large submatrix (> 2 GB) without any errors, circumventing the bug in the Intel MPI library.

### 3.5. *Parallelization of the JOREK subroutines*

After the JOREK parallel input was successfully developed and tested the rest of the code that uses the distributed matrices from STARWALL had to be parallelized as well. This mainly concerns the parallelization of the linear algebra operations.

#### 3.5.1. **Data structure of distributed matrices**

A new data structure (Fig. 28) was introduced in JOREK in order to encapsulate properties of a distributed matrix. *loc_mat* represents the local chunk (two dimensional array) of a distributed matrix. *distrib* tells us if a matrix is distributed or not. *row_wise* shows the type of the distribution: if *row_wise=.true.* the matrix is distributed row–wise; if *row_wise=.false.* the matrix is distributed column–wise. *ind_start* and *ind_end* denote the starting and ending indices of the current chunk in the global matrix. *step* is the chunk size of the local matrix. *dim* defines the global matrix dimensions.

```
type :: t_distrib_mat
    real*8, allocatable :: loc_mat(:,:)
    logical             :: distrib
    logical             :: row_wise
    integer             :: ind_start
    integer             :: ind_end
    integer             :: step
    integer             :: dim(2)
end type t_distrib_mat
```

**Fig.** 28 Data structure for a distributed matrix.

### 3.5.2. Parallelization of the *update_response* subroutine

This subroutine constitutes most of the linear algebra calculations in the code that had to be parallelized. The most time consuming operation, which appears in many places in the code, is a generalized matrix-matrix multiplication. This operation needs to be done for different combinations of distributed matrices. For clarification we take the following example from the code:

*response_m_e(:,:) = sr%a_ee(:,:) + matmul( sr%a_ey(:,:), response_m_a(:,:) ).*

Four matrices are used in this example: *response_m_e, a_ee, a_ey* and *response_m_a.* A matrix-matrix multiplication is performed between the *a_ey* and *response_m_a* matrices by using the standard sequential *matmul* subroutine. The resulting matrix is added to the matrix *a_ee* and finally saved as the *response_m_e* matrix. The difficulty is that the matrices can be distributed using different patterns (row–wise or column–wise). In addition, some small matrices in the code stay unmodified (they are not distributed). In our example, the matrices *a_ee* and *a_ey* are distributed via a row–wise pattern, the matrix *response_m_a* is distributed via a column–wise pattern and the matrix *response_m_e* is not distributed at all. There are many matrix-matrix multiplications in the JOREK code and their participating matrices have different distributions and different orders. Therefore, a suitable parallel matrix-matrix multiplication subroutine is required covering the whole spectrum of distributed (column– or row–wise) or non distributed matrices. Such a subroutine named *matrix_multiplication* was successfully developed. The subroutine works for all types of distributed matrices and generates identical results in comparison with the original code version.

Finally, the complete *update_response* subroutine was parallelized including the matrix-matrix operations as well as the matrix-vector calculations and matrix reassignments.

### 3.5.3. Parallelization of the remaining part of the JOREK code

The JOREK subroutines that include distributed matrices were parallelized next. We will not report in detail about each subroutine, because the parallelization procedure and the type of parallelization were very similar. The main difficulty was with sequential subroutines that were only called by the master MPI task. In the parallel version all tasks must call and enter these subroutines and corresponding changes for a correct execution were performed.

Here is a list of all subroutines which were parallelized: *get_vacuum_response, read_starwall_response, broadcast_starwall_response, log_starwall_response, update_response, coil_current_source, evolve_wall_currents, reconstruct_triangle_potentials, equilibrium, poisson, boundary_check, vacuum_equil, vacuum_boundary_integral.*

The accuracy of the modified subroutines was compared with the accuracy of the subroutines in the original code version. Both code versions (original and parallel) provide the same results with a relative error of $10^{-11}$.

The parallel code version was also successfully tested for a production run using input matrices from STARWALL with a size of around 500 GB.

### 3.5.4. OpenMP parallelization of the matrix multiplication subroutine

During the performance tests described below it became clear that the matrix multiplication subroutine takes most of the computational time inside the *read_starwall_response* and *update_response* subroutines. Therefore, it was decided to implement an OpenMP parallelization on top of the MPI parallelization.

Results obtained from the new MPI+OpenMP subroutine were compared with the old MPI version. They were identical up to a relative error of ~$10^{-12}$. Afterwards the execution time of the multiplication of the two largest matrices in the code was measured. This test had the following parameters: *n_tor*=11, *n_period*=1, *n_plane*=32, *n_har*=6, *n_pol*=160, *nwu*=*nwv*=300 and *ntri_w*=180000. Using 18 computing nodes and 2 MPI tasks per node with 24 OpenMP threads the wall clock time was 6.6 s. This is 16.7 times faster in comparison to the old MPI version (110.7 s).

### 3.6. *Bugs in the original code version*

During the parallelization of the JOREK code a few bugs were found in the original code version and reported to the project coordinator. Among them use of uninitialized variables and wrong parameters in subroutines:

1) The variables *heat_src* and *part_src* in the file *diagnostic/integrals.f90* were without initialization for certain conditions.
2) In the file *vacuum/vacuum_response.f90* the variables: *old_thick*, *old_res*, *old_tstep*, *old_theta*, *old_zeta*, *old_reswall* were used without initialization. The main difficulty was that all these variables have the "*save*" attribute and the standard Intel Fortran debugging flag (-check uninit) can't detect them.
3) The variable *vertical_FB* in the file models/equilibrium.f90 was also used without initialization for certain conditions.
4) A wrong parameter was used in the subroutine *integrals*. Instead of using the variable *psi_bnd* the variable *psi_lim* was used. This caused wrong output results for one particular diagnostics of the JOREK code.

A few bugs were also found in the JOREK regression tests:

1) In the file *diagnostics/rst_hdf52bin.f90* and *diagnostics/rst_bin2hdf5.f90* a call to the initialization subroutine *update_time_evol_params* was missing.
2) One minor bug is still not resolved in the regression test named *freebound_equil_aug*: "forrtl: error (65): floating invalid" appears for the following line: write(11,'(8e16.8)') surface_list%psi_values(i), dp_int/sum_dl, zjz_int/sum_dl, F0 * q / (2.d0 * PI). The project coordinator will resolve this issue after the current project.

### 3.7. *Merging different JOREK development branches*

At the end of the project three branches of the JOREK code: (i) *develop* – the main branch, (ii) *feature/IMAS-668* – the branch for the current project and (iii) *feature/IMAS-961-speed-up-boundary-int* – the branch for speeding up the two most time consuming subroutines (implemented by the project coordinator) were merged. The resulting code version was tested for accuracy and performance (results are presented in the next section). Finally, a pull request on the *git* system was initiated in order to assign this code version as the main develop version.

### 3.8. *Performance tests*

In this section we compare the performance of the most important JOREK subroutines for three code versions: (i) *develop* – the main version; (ii) the *speed up* version described in Sec. 3.7 and (iii) the *merge* of *develop*, *speed up* and the version for the current project (JORSTAR2). Table 1 shows the execution time of three different test cases for these three code versions.

One can see that the total wall clock time (last column) from the original *develop* version is much higher (depending on the test case) in comparison to the *speed up* and *merge* versions. For test case number three the *develop* version did not even finish within 24 hours of computation. On the other hand the *speed up* and *merge* versions provide very similar results for the test cases one and two. For test case number one the *speed up* version is a little bit faster in comparison to the *merge* version, while for test case number two the *merge* version takes the lead. Only the *merge* version is able to run very large problem sizes due to the STARWALL response matrices being distributed over all MPI tasks.

The performance improvement was mainly achieved in two subroutines named *vacuum_boundary_integral* and *global_matrix_structure_vacuum*. The computational algorithm was changed for the latter subroutine in a way which prevents an execution of the subroutine inside a loop. Therefore, it is called only once in the *merge* and the *speed up* version and 23 times in the *develop* version. One can also see the important improvement of the *vacuum_boundary_integral* subroutine, where the wall clock time of the *develop* version is 200–770 times higher in comparison to the wall clock time of the *merge* and the *speed up* version. This improvement was mainly achieved by reordering some nested loops (there are a total of 12 nested loops) inside an OpenMP region.

The *boundary_check* subroutine was in some cases slower in the *merge* version in comparison to the *develop* and/or *speed up* version. Therefore, it was decided to implement an OpenMP parallelization for it as well. The project coordinator was responsible for this part. The subroutine after the improvement is about one order of magnitude faster using 48 OpenMP tasks in comparison to the original version.

The *merge* code version works as fast as the *speed up* version and much faster than the original *develop* version. Moreover, the *merge* version can perform calculations with larger matrices due to the MPI parallelization and a global matrix distribution of the *vacuum_response* part of the JOREK code. This is the reason why test case number three fails in the *develop* and *speed up* version due to memory limitations. On the other hand, the *merge* version executes this test case without any problems.

| 1) Test case: n_tor=21, n_period=1, n_plane=64, n_har=11, n_pol=120, nwu=nwv=64, MPI=11, compute_nodes=11, OMP=48 | | | | | |
|---|---|---|---|---|---|
| Code version | vacuum boundary integral | boundary check | update response | global_matrix structure vacuum | complete code |
| *develop* | 2329,27 | 3,44 | 21,55 | 8,74*23 | **54903** |
| *speed up* | 5,89 | 1,04 | 21,01 | 0,62 | **1177** |
| *merge* | 4,57 | 4,63 | 12,49 | 0,60 | **1308** |
| 2) Test case: n_tor=21, n_period=1, n_plane=64, n_har=11, n_pol=160, nwu=nwv=64, MPI=11, compute_nodes=11, OMP=48 | | | | | |
| Code version | vacuum boundary integral | boundary check | update response | global_matrix structure vacuum | complete code |
| *develop* | 11147,05 | 8,74 | 92,05 | 19,46*10 | **>24 hours** |
| *speed up* | 14,45 | 2,53 | 37,81 | 1,19 | **1883** |
| *merge* | 9,91 | 6,20 | 22,54 | 1,03 | **1649** |
| 3) Test case: n_tor=11, n_period=1, n_plane=32, n_har=6, n_pol=160, nwu=nwv=330, MPI=48, compute_nodes=24, OMP=24 | | | | | |
| Code version | vacuum boundary integral | boundary check | update response | global_matrix structure vacuum | complete code |
| *develop* | Not enough memory | | | | |
| *speed up* | Not enough memory | | | | |
| *merge* | 1,35 | 9,52 | 213,80 | 0,29 | **2385,00** |
| All values are given in seconds | | | | | |

**Table** 1. The wall clock time in seconds of the most important JOREK subroutines for three different test cases for three code versions.

### 3.9. *Scalability tests*

We also tested how the computational time scales for the final *merge* version for different parameters of the code. Fig. 29 shows the wall clock time versus (a) the number of finite element triangles representing the wall, (b) versus the number of MPI tasks and (c) versus the number of the nodes in the poloidal direction in the JOREK grid. In test (a) we kept the number of MPI tasks=12 and *n_pol*=200 constant; in test (b) we set *ntri_w*=100,000 and *n_pol*=200 and in test (c) we used MPI tasks=12 and *n_pol* =200.

The total computational time for all test cases (even for production runs with *ntri_w*=300,000 or *n_pol*=300) and for any presented subroutine is not longer than 70 seconds. These results are satisfactory as all these subroutines, except for the *boundary_check,* are called only once. The *boundary_check* subroutine is executed inside a small loop, however the calculation time of this subroutine is less than two seconds for a large production run which makes its influence on the total computational time relatively small. Besides, the parallelized *read_starwall_response* and *matrix-matrix* multiplication subroutine scale quite well. All obtained data, which is used for Fig. 29, is summarized in Table 2.

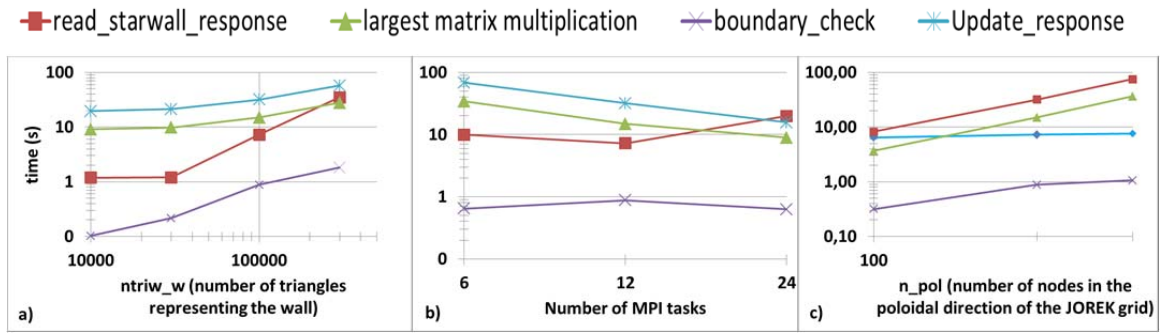

**Fig.** 29 Computational time in seconds of some JOREK subroutines versus different parameters of the code. The following constant parameters were used in all calculations: *n_tor*=11, *n_period*=1, *n_plane*=32, *n_har*=6.

| MPI | ntri_w | n_pol | read starwall response | update response | Matrix–matrix multiplication | boundary check |
|---|---|---|---|---|---|---|
| 12 | 10,000 | 200 | 1,19 | 19,65 | 9,06 | 0,10 |
| 12 | 30,000 | 200 | 1,20 | 21,19 | 9,74 | 0,22 |
| 12 | 100,000 | 200 | 7,24 | 31,77 | 14,88 | 0,88 |
| 12 | 300,000 | 200 | 35,03 | 57,33 | 27,72 | 1,81 |
| 6 | 100,000 | 200 | 9,96 | 68,47 | 34,32 | 0,64 |
| 24 | 100,000 | 200 | 19,93 | 15,68 | 8,91 | 0,62 |
| 12 | 100,000 | 100 | 6,47 | 8,18 | 3,68 | 0,32 |
| 12 | 100,000 | 300 | 7,64 | 75,20 | 36,32 | 1,06 |

**Table** 2 Computational time in seconds of the four parallelized subroutines in the JOREK-STARWALL part.

The execution time of one time step in the global loop was measured in the JOREK code with and without the STARWALL part in order to estimate the overhead of this part of the code (Fig. 30). The time of the JOREK-STARWALL run is about a factor of two higher in comparison to the pure JOREK run. It grows with increasing problem size, while the JOREK part stays constant (a). This is because the tested parameter (*ntri_w*) has no influence on the JOREK part.

There are five nested subroutines in the JOREK-STARWALL part: *construct_matrix* → *vacuum_baundary_integral* → *evolve_wall_currents* → *write_wall_vtk* → *reconstruct_triangle_potentials*. It was measured that the time difference between the pure JOREK and the JOREK-STARWALL code version came mainly from the last two subroutines *reconstruct_triangle_potentials* and *write_wall_vtk*. For example, the execution time of the *write_wall_vtk* subroutine for the test case Fig. 30 (a) with *ntri_w*=300,000 is 21 seconds. In comparison the kernel loop of the *vacuum_baundary_integral* subroutine for this test case takes about two seconds. Thus, *write_wall_vtk* can be the first candidate for a future code optimization. All obtained data, which is used for Fig. 30, is summarized also in Table 3.

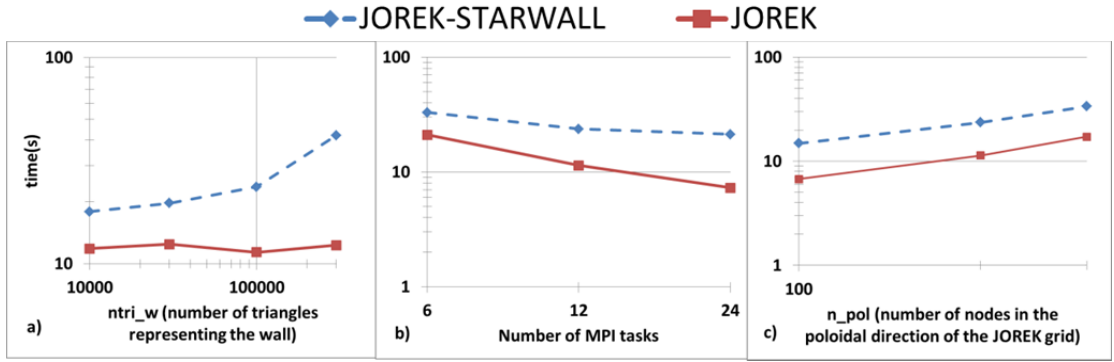

**Fig.** 30 Computational time in seconds of one global loop step in the JOREK code with the STARWALL part (dashed blue line) and without (solid red line) versus different parameters of the code. The following constant parameters were used in all calculations: $n\_tor$=11, $n\_period$=1, $n\_plane$=32, $n\_har$=6.

| MPI | ntri_w | n_pol | JOREK | JOREK–STARWALL |
|---|---|---|---|---|
| 12 | 10,000 | 200 | 11,85 | 17,88 |
| 12 | 30,000 | 200 | 12,42 | 19,69 |
| 12 | 100,000 | 200 | 11,37 | 23,60 |
| 12 | 300,000 | 200 | 12,29 | 42,00 |
| 6 | 100,000 | 200 | 21,05 | 32,92 |
| 24 | 100,000 | 200 | 7,27 | 21,27 |
| 12 | 100,000 | 100 | 6,75 | 14,85 |
| 12 | 100,000 | 300 | 17,17 | 33,60 |

**Table** 3 Computational time in seconds of one global loop step in the pure JOREK and JOREK–STARWALL version.

Additionally, a prediction for the memory consumption (total and per MPI task) was implemented in JOREK and STARWALL to help the user to choose the appropriate number of MPI tasks.

## 4. Conclusions

The STARWALL code has been analyzed for potential improvements and optimization by means of MPI parallel computation. It was found that for a large production run the whole code must be parallelized due to the lack of memory for saving the input/output matrices and due to the computational time.

All sequential LAPACK subroutines were analyzed and selected for replacement by their parallel analogues from the ScaLAPACK library. All these subroutines were replaced in the final code version because of the required large input matrices size.

The LAPACK subroutine for the eigenvector solver was replaced by the parallel subroutine counterpart from the ScaLAPACK library. A very good agreement was found in terms of the eigenvalues. In addition, the correctness of the results was proven by their consistency with the underlying physical model. The ScaLAPACK subroutine has shown better performance not only by using several processes in parallel but also in sequential mode due to the advantage of using IEEE arithmetics (optimization of arithmetic operation with $\pm\infty$) [9, page 121]. Finally, good parallelization efficiency was obtained for this subroutine for large problem sizes.

The subroutines *matrix_ww*, *matrix_pp, matrix_wp* and *tri_induct* were re-written in order to avoid the storage of the largest matrices in the code named *dima* and similar. This allows to save significant fraction of the memory that will bring the opportunity to perform calculations for larger problem sizes. The subroutines were parallelized with MPI taking into account the specific output index format for matrices which is necessary for ScaLAPACK subroutines. A good scalability was achieved for all subroutines with a speed-up factor of more than 210 when 512 cores were involved in the computation.

Finally, the complete code was parallelized including all LAPACK and user written subroutines. The new parallel version of the code provides identical results in comparison with the original code. This includes the part of the code handling the magnetic coils. The parallelized version allows production runs with much larger numbers of finite elements that allows to resolve realistic wall structure. The simulation time in such a case is less then 12 hours using 128 computing nodes on HELIOS.

Different libraries (e.g. ScaLAPACK, HDF5 and MPI) were analyzed in order to find the best possible solution for parallel I/O. The MPI library was chosen as it can directly translate the format of the output submatrices from the block-cycling distribution to an ordinary format during the writing procedure.

MPI parallel I/O was implemented in both the STARWALL output and the JOREK input subroutines. The execution time of the complete reading and writing procedure for a production run is only up to a few minutes when using 16 and 64 computing nodes.

During the development, a bug was found in the Intel MPI library that significantly limits the size of the read/written data per operation. It was reported to the Intel support team and should be corrected within the next version of the library. In order to be able to work with large matrices and to overcome this bug, several subroutines were modified using a workaround.

All sequential subroutines in JOREK that use distributed matrices from the STARWALL input file were parallelized and tested. All of them provide identical results in comparison to the original code version.

The final version of the code, achieved during this project, was merged with the *develop* branch and with the *speed up* version of the code (where loops had been re-ordered in order to speed up the code). The final code version provides results much faster than the *develop* version and can work with very large matrices from STARWALL output.

## Acknowledgment

This work has been carried out within the framework of the EUROfusion Consortium and has received funding from the Euratom research and training programme 2014-2018 under grant agreement No 633053. The views and opinions expressed herein do not necessarily reflect those of the European Commission. Part of this work was carried out using the HELIOS supercomputer system at the Computational Simulation Centre of the International Fusion Energy Research Centre (IFERC-CSC), Aomori, Japan and the Marconi supercomputer system at the CINECA research center, Bologna, Italy.